# Effect of nanometer-sized B powder on phase formation of polycrystalline $MgB_2$

Ling An [1,3], Chinping Chen [1], Bo Wang [2], Cheng-gang Zhuang[1], Xing-guo Li[2], Zeng-jun Zhou[1], and Qing-rong Feng[1*]

1. Department of Physics and State Key Laboratory of Artificial Microstructure and Mesoscopic Physics, Peking University, Beijing 100871, P.R. China
2. Department of Chemistry, Peking University, Beijing, 100871, P.R. China
3. Department of Physics, Chang Ji Collage, Xinjiang, 831100, P.R. China

## Abstract

The size effect of the raw B powder on the $MgB_2$ phase formation has been studied by the technique of *in-situ* high temperature resistivity (HT-ρT) measurement. The onset temperature, $T_{onset}$, and the completion temperature, $T_{PF}$, of the phase formation are determined directly during the ongoing thermal process. These two temperatures, $T_{onset}$ and $T_{PF}$ of the sample synthesized using nanometer B and Mg powders (NanoB-$MgB_2$) are 440 °C and 490 °C, respectively, the same as those of the sample using micrometer B and nanometer Mg powders (MicroB-$MgB_2$). This indicates that the phase formation temperature of $MgB_2$ do not depend on the B powder size. On the other hand, the upper limit of the sintering temperature, $T_N$, above which the sample loses superconductivity, is below 750 °C for NanoB-$MgB_2$, much lower than 980 °C for the $MgB_2$ prepared using micron-sized B powder and millimeter sized Mg powder (DM-$MgB_2$). In comparison with the sample directly sintered at 650 °C < $T_N$, an interesting, irreversible transformation in the crystal structure of the $MgB_2$ phase was observed with the sample going through the stages of initial sintering at 750 °C, then re-sintering at 650 °C in an Mg-rich environment after the processes of regrinding and pressing. Possible explanation of the observed properties is discussed.

Corresponding author: Qing-rong Feng,
Tel : +86-10-62751731, Fax: +86-10-62751615, E-mail: qrfeng@pku.edu.cn

## Introduction

The superconductivity of $MgB_2$ at 39 K has stimulated a lot of research activities in the field of superconducting science. In the past few years, great advances have been made in the understandings of the physical properties[1] and in the recipes of the synthesis techniques. The transition temperature at 39 K makes the cryo-cooler a convenient apparatus to provide the necessary cryogenics environment without the trouble of handling liquid helium. In addition, the upper critical field, $H_{C2}$, is reported to reach as high as 50 T[2] and the critical current density $J_C > 10^7$A/cm$^2$. These properties indicate that $MgB_2$ is promising to play an important role in the next generation superconducting industry. Therefore, a deeper insight into the synthesis technique for the various forms of the sample, such as polycrystalline[3,4], film[5,6], wire[7-9], tape, *etc.*, is significant.

The fabrication technique for the polycrystalline samples has already been widely investigated. The pressure and the temperature of the gas environment in the furnace and the duration of the thermal process inside the furnace are the main parameters often explored for a recipe of an optimum process[10]. On the other hand, the effect of the raw materials, especially of the powder size, on the $MgB_2$ phase formation is another important parameter worth studying. In the previous reports, the evolution of the phase formation and crystal structure at various furnace temperature[3] as well as the dependence of the phase formation temperature, $T_{PF}$, on the raw Mg powder size[4] have been studied by the *in situ* HT-ρT technique. This has added a new dimension to the knowledge for the production of a better quality polycrystalline $MgB_2$. In the present work, the effect of the B powder size on the characteristic thermal temperature, $T_{onset}$ and $T_{PF}$, and the optimum sintering temperature limit to form superconducting $MgB_2$ phase has been investigated. Both of the raw B and Mg powders are in the nanometer scale. The $MgB_2$ crystal phase were studied using a Philip x' pert diffractometer, using the Cu $K_\alpha$ light source. The Meissner state is determined by a M-T measurement performed on a Quantum Design MPMS SQUID magnetometer. The result of the phase formation for NanoB-$MgB_2$ is compared with that of the MicroB-$MgB_2$ and DM-$MgB_2$ reported previously[4,10].

## Experimental

NanoB-$MgB_2$ samples were prepared by the solid-state reaction method using the raw materials of nanometer boron powder (99%) ~ 100 nm, and nanometer magnesium powder, ~ 40 nm. The nanometer sized B and Mg powders were produced in the laboratory by the method of arc discharge. These raw powders were mixed thoroughly and pressurized into a mould made of $MgB_2$ embryo crystals. The resistivity of the sample inside the mould was monitored by a standard four-probe method using the Pt wires, ϕ 0.4 mm, as the electrical leads. It was then put into the furnace in the argon atmosphere at ambient pressure for the heat treatment. The temperature condition of the heat treatment was similar to that in the fabrication of MicroB-$MgB_2$[3]. For a complete thermal cycle, the temperature was initially raised to the sintering temperature, for example 750 °C, at a rate of 340 °C per hour, then, held constant for 40 minutes. In the end, the sample was cooled down to room temperature inside the furnace without any regulation.

## Results and Discussion

The HT-ρT result of the NanoB-$MgB_2$ is presented in Fig. 1, the solid circles, in a logarithmic vertical scale. For comparison, the result of MicroB-$MgB_2$ is also plotted in a linear vertical scale. Both results exhibit similar features in the variation of resistivity during the thermal process. The onset of the $MgB_2$ phase occurs, in both cases, at $T_{onset}$ ~ 440 °C, corresponding to the maximum in the HT-ρT curve. At $T < T_{onset}$, the resistivity increases with the increasing temperature for both samples, reflecting the metallic property of the Mg nanoparticles surrounding the larger B particles. This indicates that the $MgB_2$ phase formation does not take place yet. At $T > T_{onset}$, on the other hand, the resistivity goes down rapidly till a characteristic temperature, $T_{PF}$, is reached. The abrupt drops of the resistivity within the temperature window, $\Delta T_{PF} = T_{PF} - T_{onset}$, is a sign for the $MgB_2$ phase formation. The differential of the HT-ρT curve with respect to the temperature, dρ/dT, is plotted in Fig. 2. Large fluctuation exists in dρ/dT at $T < T_{onset}$, before the appearance of the $MgB_2$ phase. The section in which dρ/dT < 0, corresponding to the steepest decreasing section in Fig. 1, clearly demonstrates the temperature window, $\Delta T_{PF}$. For NanoB-$MgB_2$, $\Delta T_{PF}$ ~ 55 °C, and MicroB-$MgB_2$, 50 °C. The above results demonstrate that $T_{PF}$ and $\Delta T_{PF}$ are almost the same for both of the samples, suggesting that the particle size of the B powder is not a crucial factor in the determination of the $MgB_2$ formation temperature, as is the Mg powder size[4]. As the temperature keep going up, higher than $T_{PF}$, the differential resistivity, dρ/dT, approaches zero again with much less fluctuations.

As the temperature exceeds the upper bound, $T_N$, of the optimum temperature range for sintering, the sample loses the superconductivity. However, there is no apparent signature corresponding to $T_N$ showing up in the HT-ρT curve in Fig 1. Therefore, one has to rely on the indirect M-T measurement, probing the Meissner state similar to that presented in Fig. 3 of reference 3, in order to determine $T_N$. For NonoB-$MgB_2$, $T_N$ is thus determined below 750 °C. This is much lower than 980 °C for the $MgB_2$ sample prepared using micrometer sized B powder in the vacuum condition[4]. With the $T_N$ determined, a thermal cycle process on the NanoB-$MgB_2$ sample has been performed. Two NanoB-$MgB_2$ samples were prepared. After the initial rising in temperature, one of the samples has been sintered at 650 °C for 40 minutes (650-NanoB-$MgB_2$), the other, 750 °C for 2 hours (750-NanoB-$MgB_2$). The M-T measurements revealed that 650-NanoB-$MgB_2$ was superconducting with a very broad transition temperature, about 20 K, see Fig. 3, while 750-NanoB-$MgB_2$ did not show any sign of Meissner state. The sample, 750-NanoB-$MgB_2$, was then pulverized, reground, and pressed into a $MgB_2$ embryo box for the re-sintering at 650 °C for 40 minutes. The re-sintering process has turned the sample into superconducting state with a much sharper transition width than that of 650-NanoB-$MgB_2$, see the solid circles in Fig. 3.

The evolution of the $MgB_2$ crystal phase, resulting from the afore-mentioned thermal processes, is of interest. The study has been carried out by the XRD measurement. The XRD spectrum presented in Fig. 4 are for 650-NanoB-$MgB_2$ (labeled as A), 750-NanoB-$MgB_2$ (B), and 750-NanoB-$MgB_2$ re-sintered at 650 °C (C). The indexed peaks in the figure correspond to the $MgB_2$ phase. Apparently, the $MgB_2$ phase, existing in the spectrum A for 650-NanoB-$MgB_2$, disappears completely in the spectrum B for 750-NanoB-$MgB_2$. The impurity phases in both of the spectrum, including the Mg phase marked as α, and the MgO phase, β, are the same, except for the peak marked by an asterisk in spectrum B. The asterisked peak cannot be indexed to (101) of $MgB_2$, since not a sign of superconductivity is observed in the M-T measurement. In the spectrum C, the $MgB_2$ phase appears again for 750-NanoB-$MgB_2$ re-processed at 650 °C, but with a transformed $MgB_2$ phase different from that of 650-NanoB-$MgB_2$. On the other hand, the MgO phase, marked as β, exists for all of the samples, indicating that it is relatively stable. It, therefore, becomes a benchmark for the other crystal phase to compare with.

It is interesting to note that the asterisked peak in the spectrum B cannot be reasonably indexed to any of the $MgB_2$, Mg, or MgO phases at all. One of the most likely scenarios is that the high sintering temperature at 750 °C makes the Mg atoms to escape, leaving behind vacancies in the crystal structure. As a result, the superconductivity is destroyed. The asterisked peak is therefore corresponding to the Mg-deficient (101) peak in the spectrum A, with the diffraction peak position shifted up by about 0.3 degree. Accordingly, the lattice constant is reduced by about $7\times10^{-4}$ nm, estimated using the Bragg diffraction equation. With the reground, and re-sintering processes of the 750-NanoB-$MgB_2$ at 650 °C in an Mg-rich environment, the superconductivity shows up again, as is evidenced by the Meissner state in the M-T measurement, Fig. 3, and the XRD peaks in the spectrum C of Fig. 4. Interestingly, the major $MgB_2$ peaks in spectrum C are quite different from those appearing in spectrum A. It seems that the re-introduction of Mg atoms into the crystal vacancies does not restore the modified lattice constant resulting from the Mg escaping. This is apparent by the fact that the (101) peak in the spectrum C occurs at the same position as that of the asterisked peak in the spectrum B, but differs from the (101) peak in the spectrum A.

## Conclusion

The effect of the B powder size on the polycrystalline $MgB_2$ phase formation has been investigated. According to the HT-ρT measurement, the onset and completion temperatures, $T_{onset}$ and $T_{PF}$, of the $MgB_2$ phase formation are almost unaffected by the variation of the B powder size. However, the upper limit, $T_N$, above which the sample loses superconductivity, is considerably lowered at the reduced B powder size. It decreases from 980 °C with DM-$MgB_2$ prepared using micron-sized B powder and millimeter sized Mg powder [10], down to below 750 °C for the one prepared using nanometer sized B powder. An interesting transformation in the crystal structure of the $MgB_2$ phase has occurred as shown in the spectrum A and C in Fig. 4, accompanying by a variation of lattice constant clearly revealed by the (101) peak position. This may possibly due to the escape and re-entry of the Mg atoms into the $MgB_2$ phase during the thermal processes.

## Acknowledgement
We wish to thank Mr. Jun Xu and Yong-zhong Wang for technical assistance. This work was support in part by the Department of Physics of Peking University, and in part by the Center for Research and Development of Superconductivity in China under contract No. BKBRSF-G1999064602. L. An visited the School of Physics of Peking University from March 2004 to July 2004.

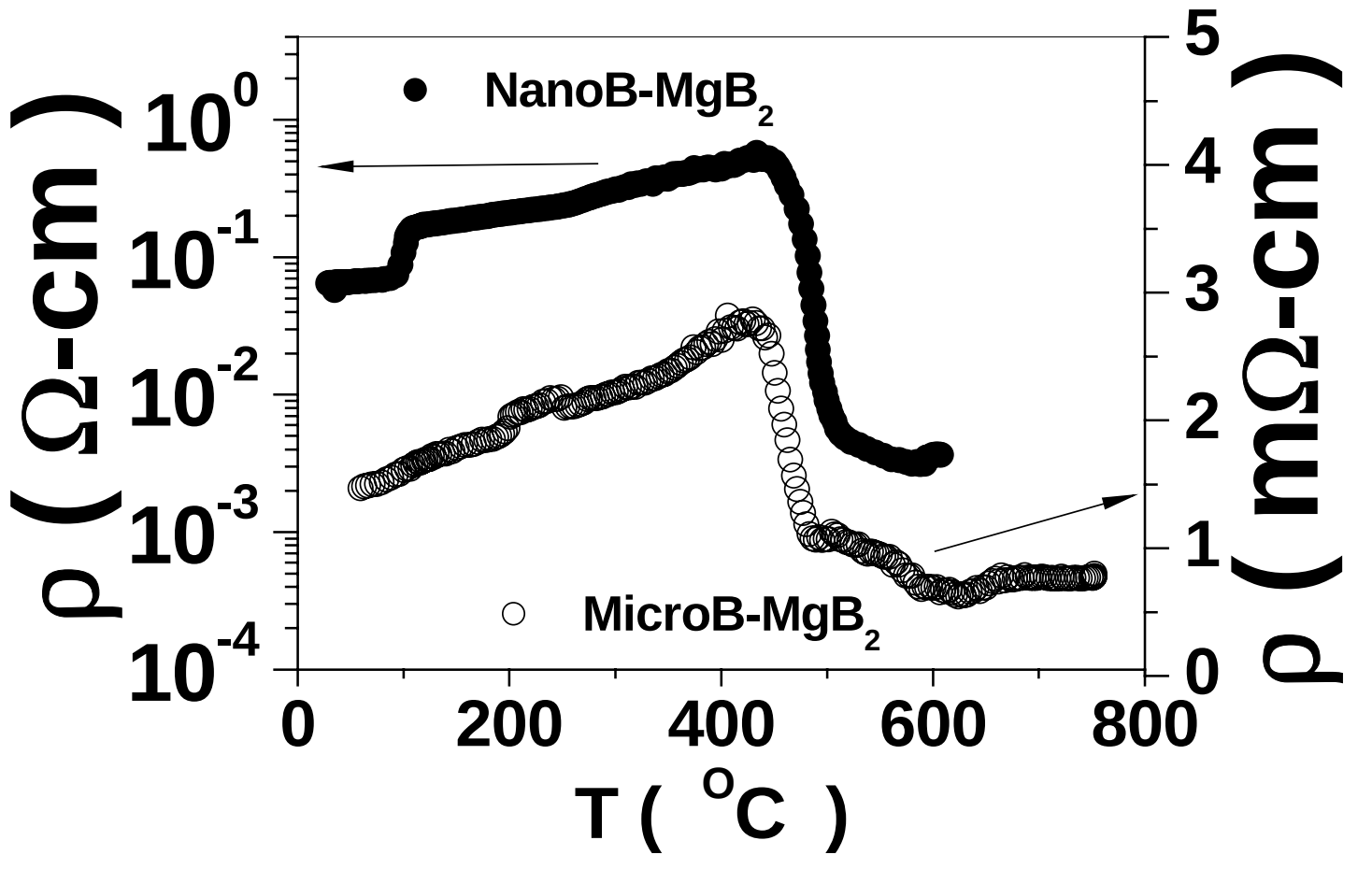

Fig.1. *In-situ* HT-ρT curves for NanoB-$MgB_2$ and MicroB-$MgB_2$[3]. The signature for the phase formation temperature, $T_{onset}$, occurs at the same temperature, 440 °C, for both of the samples.

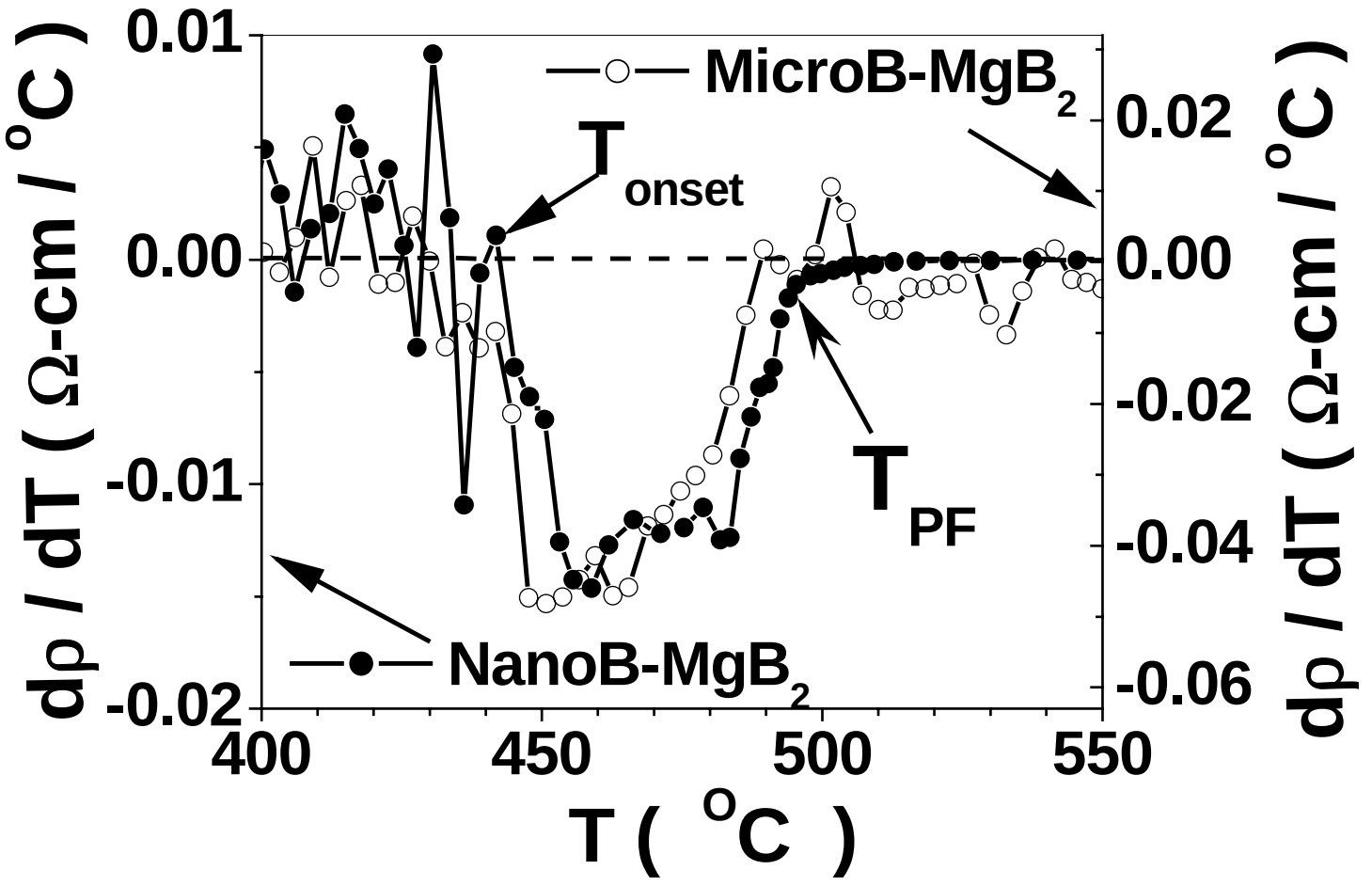

Fig. 2 Differential resistivity, dρ/dT, derived from the HT-ρT data. The onset and completion of phase formation temperature, $T_{onset}$ and $T_{PF}$, are the same, as clearly indicated in the figure.

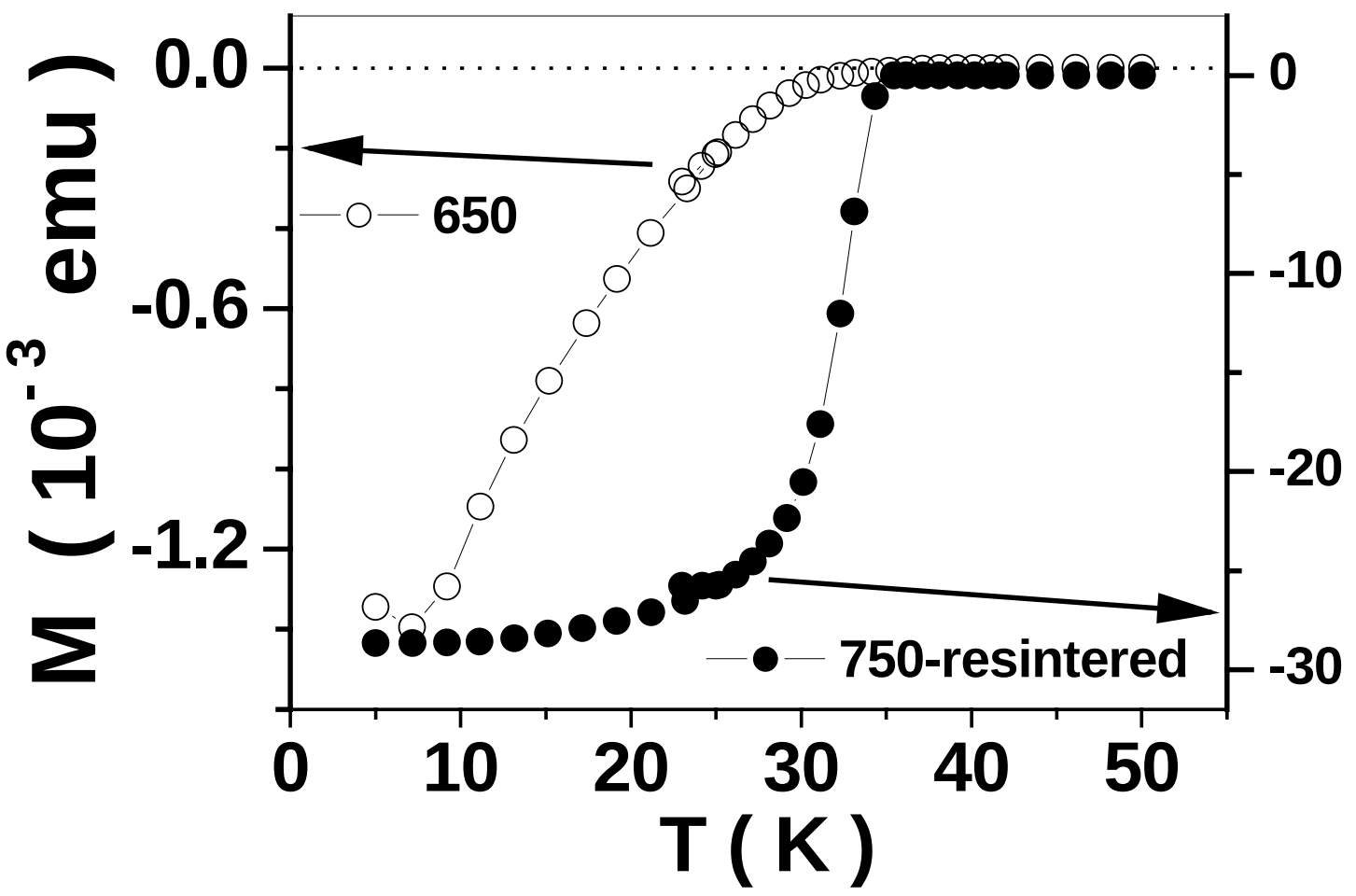

Fig. 3 M-T of 650-NanoB-$MgB_2$, open circle, and 750-resintered NanoB-$MgB_2$, slid ciecle, under an applied field of 100 Oe. The 650-NanoB-$MgB_2$ sample has a very broad transition width, about 20 K, while the 750-resintered one exhibits a much sharper transition width of about 5K.

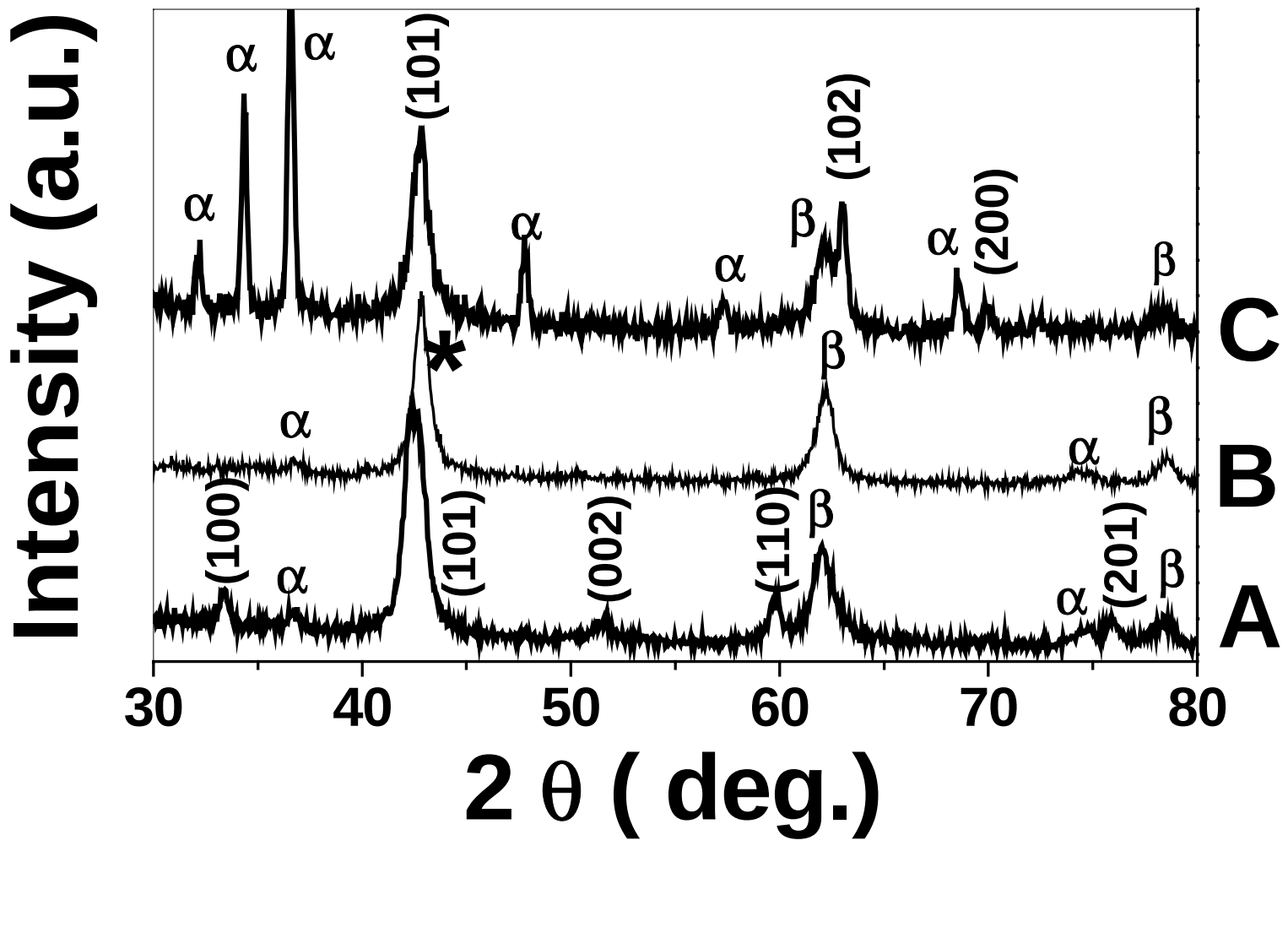

Fig 4. XRD spectrum for 650-NanoB-$MgB_2$ (labeled as A); 750- NanoB-$MgB_2$ (B), and 750- NanoB-$MgB_2$ re-sintered at 650℃ (C). The zero positions in the vertical scale of the spectrum B and C are offset to plot in the same diagram with spectrum A. The indexed peaks are for the $MgB_2$ phase. The peaks labeled as $\alpha$ are for the Mg phase and $\beta$, MgO while the one marked by the asterisk in spectrum B remains open for discussion.